\documentclass[%
 reprint,
 amsmath,amssymb,
 aps,
]{revtex4-1}

\usepackage{graphicx}% Include figure files
\usepackage{dcolumn}% Align table columns on decimal point
\usepackage{bm}% bold math
\begin{document}

\preprint{APS/123-QED}

\title{Intense source of cold cesium atoms based on a two-dimensional magneto-optical trap with independent axial cooling and pushing}% Force line breaks with \\

\author{J. Q. Huang$^{1,3}$, X. S. Yan$^{2,3}$, C. F. Wu$^{1,3}$, J. W. Zhang$^{2,3}$, Y. Y. Feng$^{2,3}$}
% \altaffiliation[Also at ]{Physics Department, Tsinghua University.}%Lines break automatically or can be forced with \\
\author{L. J. Wang$^{1,2,3,}$}%
 \email{lwan@tsinghua.edu.cn}

\affiliation{%
 $^{1}\mbox{Department of Physics, Tsinghua University, Beijing 100084, P. R. China}$\\
 $^{2}\mbox{Department of Precision Instrument, Tsinghua University, Beijing 100084, P. R. China}$\\
 $^{3}\mbox{Joint Institute for Measurement Science (JMI), Tsinghua University, Beijing 100084, P. R. China}$\\
}%


% \homepage{http://www.Second.institution.edu/~Charlie.Author}%
%\affiliation{
%$^{2}Department of Precision Instrument, Tsinghua University, Beijing 100084, P. R. China$
%\affiliation{
%\affiliation{%


\date{\today}% It is always \today, today,

\begin{abstract}
We report our studies on an intense source of cold cesium atoms based on a two-dimensional magneto-optical trap with independent axial cooling and pushing. The new-designed source, proposed as 2D-HP MOT, uses hollow laser beams for axial cooling and a thin pushing laser beam for cold atomic beam extraction. Regulated independently by the pushing beam, the atomic flux can be substantially optimized. The atomic flux maximum obtained in the 2D-HP MOT is $4.02\times 10^{10}$ atoms/s, increased by 60 percent compared to the traditional 2D$^+$ MOT in our experiment. Moreover, with the pushing power 10 $\mu$W and detuning $0\Gamma$, the 2D-HP MOT can generate a rather intense cold cesium atomic beam with the concomitant light shift suppressed by 20 times in magnitude. The axial velocity distribution of the cold cesium beams centers at 6.8 m/s with a FMHW of about 2.8 m/s. The dependences of the atomic flux on the pushing power and detuning are studied. The experimental results are in good agreement with the theoretical model. 
\begin{description}
\item[PACS numbers]
37.10.Gh, 37.20.+j, 37.10.De
\end{description}
\end{abstract}

\pacs{Valid PACS appear here}% PACS, the Physics and Astronomy
\maketitle

%\tableofcontents

\section{\label{sec:level1}INTRODUCTION }

Cold atomic beams are widely used in fields of atom interferometry \cite{1Kasevich,1muller,2xue2015}, atomic clocks \cite{3Gibble,4Dimarcq,5HWang,5j.vanier}, cold atom collisions \cite{3Gibble,6J.Weiner,7Schollkopf} and atom optics\cite{8Adams}. They are also required for fast loading a magneto-optical trap (MOT) in ultra high vacuum (UHV). The desirable features of cold atomic beams are high flux at low mean velocity, narrow velocity distribution, and small divergence.   

The cold atomic beams can be produced by using a Zeeman slower \cite{9F.Lison} or an isotropic-light tube \cite{91isotropic} to decelerate a thermal atomic beam along its propagation axis, or by extracting the cold atoms from a vapor MOT directionally. Compared to the decelerator, the MOT source is preferred because of its narrower velocity distribution and higher brightness. The MOT source can be implemented in various methods, including the low-velocity intense source (LVIS) \cite{2xue2015,10Z.T.Lu}, the moving molasses (MM) MOT \cite{11Thomann}, the pure two-dimensional (2D) MOT \cite{12J.Schoser}, the 2D MOT combined with a thin pushing beam \cite{15J.Catani} and the 2D$^{+}$ MOT \cite{13K.Diechmann}. 

The 2D$^{+}$ MOT source is a configuration in which the two-dimensional magneto-optical trap is complemented with a pair of laser beams in the axial direction for optical molasses cooling. The axial unbalanced radiation pressure caused by a centric aperture pushes the cold atoms out of the MOT. Equipped with both cooling and pushing along the axial direction, the 2D$^{+}$ MOT have a higher efficiency than other configurations of the MOT source. And to obtain higher atomic flux, the axial laser beams were optimized in various ways, including the power and detuning \cite{14L.Cacciapuoti,16S.Chaudhuri}, the composition \cite{17S.J.Park}, and the structure \cite{18Fang}. But in these optimizations, the axial laser beams are used for cooling and pushing without distinction. Considering the physical process in the MOT, cooling decelerates atoms, but pushing accelerates atoms. For optimal performance, the pushing actually requires different laser power and detuning from the cooling. The 2D$^{+}$ MOT cannot be substantially optimized until the cooling and pushing are regulated separatedly. 

To achieve further improvement for the 2D$^{+}$ MOT, we propose a novel configuration for the axial laser beams. In the scheme, a pair of counterpropagating hollow laser beams are applied for axial molasses cooling. Another homocentric thin laser beam is used for pushing. The hollow beam and the pushing beam are functionally seperated and make no interference to each other.

One major benefit of this configuration is that the atomic flux can be enhanced substantially. While the hollow beams for the axial cooling optimize the cold atom loading, the pushing laser beam with appropriate power and detuning can maximize atomic flux \cite{23Pillet}. The maximized atomic flux can not only help reduce the loading time of a UHV MOT, which makes great sense for precision measurements such as atomic fountain clocks \cite{Thomann_beam,fountainload}, but also greatly boost the signal-to-noise ratio (SNR) of experiments using cold atomic beam\cite{2xue2015,5j.vanier,PhysRevLett.100.180405}.

Another major benefit is the thin pushing laser leads to a significant reduction of the leaking laser light from the extracting aperture and consequently is an effective suppression on the light shift. In the traditional configuration, the axial laser beams for molasses cooling are relatively intense, which leads to relatively intense laser light leakage. The leaked light would cause considerable light shift to the cold atomic beam, which should be avoided in precision measurements, for example, in atomic clocks or atom interferometers\cite{1muller,2xue2015,5HWang,5j.vanier}. Additionally, the leaking light causes damages to subsequent optical pumping and probing. In the configuration with a thin pushing laser, the leaking laser can be reduced significantly. The light shift and other damage effects will be consequently suppressed. In addition to the two major benefits, the configuration with independent axial cooling and pushing also provides convenience for flux and velocity tuning, and beam parameter stabilization.

In this paper, we present both theoretical and experimental studies on the 2D$^{+}$ MOT with independent axial cooling and pushing. And we propose ``2D-HP MOT" to stand for this new scheme distinguishing from the traditional 2D$^{+}$ MOT, where H means the hollow cooling beams and P the thin pushing beam. The experimental results demonstrate that the atomic flux can be increased by 60 percent in the 2D-HP MOT compared to the traditional 2D$^{+}$ MOT in our experiment. Meanwhile, the calculated light shift due to the leaking laser is suppressed by 20 times in magnitude. The experimental results are in good agreement with the theoretical calculation.

This paper is organized as follows. In Sec. II, we explain the basic principle of the 2D-HP MOT, including the rate and trapping equation, the atomic flux analysis and the light shift effect. In Sec. III, we present a detail description of the experimental setup and the measurement of the cold atomic beam. The experimental results and discussions are arranged in Sec. IV. We summarize and give an outlook on future experiments in Sec. V. 

\section{PRINCIPLE OF OPERATION AND THEORATICAL MODEL}

The flux of the cold atomic beam extracted from the 2D-HP MOT is determined by the rate equation, which is the relationship between the loading rate and the loss rates. The loading rate $R$ indicates the capture capability of the MOT, which is derived from the flux through the surface of the cooling volume of atoms with a radial velocity below the capture velocity $v_r<v_c$ \cite{12J.Schoser,19C.Monroe}. Its definition per axial velocity interval $[v_z, v_z+{\rm d}v_z]$ in cylinder coordinate can be written as
\begin{eqnarray}
R(n,v_z)&=& 2\pi r_cn(\dfrac{m}{2\pi {\rm k_B}T})^{3/2}{\rm exp}(-\dfrac{mv_z^2}{2{\rm k_{B}}T})\nonumber\\
&\times&\int_0^{v_c(v_z)}v_r{\rm exp}(-\dfrac{mv_r^2}{2{\rm k_B}T})2\pi v_r{\rm d}v_r,
\end{eqnarray}
\noindent where $r_c$ is the radius of the cooling volume, $n$ the density of the thermal background vapor, $m$ the atom mass, $k_{B}$ the Boltzmann constant and $T$ the temperature of the vapor. According to Eq. (1), a high capture velocity results in a large loading rate. The capture velocity $v_c$ depends on the axial velocity $v_z$. At low $v_z$, the cooling time is so long that the capture velocity can be simply regarded as a constant $v_{c0}$, which is around 30 m/s. And at high $v_z$, the capture velocity falls off as 1/$v_z$ \cite{12J.Schoser}. In the 2D-HP MOT, the loading rate is optimized by applying the hollow laser beams to slow down atoms in the axial direction. With $v_c=30 m/s$ and $r_c=8 mm$ (see in Sec. III), the total loading rate of the 2D-HP MOT over all the longitudinal velocity is approximately $5\times 10^{10}$ atoms/s. 

The loss of the cold atoms within the MOT is mainly caused by three factors, which are the collisions between the cold atoms and the thermal background vapor (``cold-hot'' collisions), the collisions between two cold atoms (``cold-cold'' collisions), and the outcoupling from the cold atomic cloud into the beam, respectively. The loss rate due to the ``cold-hot'' collisions is proportional to the density of the thermal background vapor, which can be written as $\Gamma _{trap}=n\sigma v_{rms}$. Here $v_{rms}$ is the root-mean-square velocity of thermal atoms, and $\sigma$ is the effective collision cross section. For a cesium atom, $\sigma$ is $2\times 10^{-13} cm^2$. The loss rate caused by the ``cold-cold'' collisions is $\beta_N N/V$, growing as the density of the trapped cold atoms. Here $\beta_N$ is the trap-loss parameter, which contains the probabilities for inelastic processes, such as fine-structure-changing collisions, radiative escape and photoassociation. The study on $\beta_N$ has been an interesting and important topic. Both theoretical and experimental works verify that $\beta_N$ increases linearly with the MOT laser intensity \cite{20M.Anwar,21J.Weiner,22Pritchard}. The loss rate due to the outcoupling is called the outcoupling rate $\Gamma_{out}$, which represents the capability of the pushing laser beam extracting the cold atomic beam from the MOT. And it can be simply written as $\Gamma_{out}=1/t_{out}$, where $t_{out}$ is the time for pushing the cold atoms out of the MOT \cite{12J.Schoser,23Pillet,24Zhangtc}.

In the 2D-HP MOT, the time $t_{out}$ is determined by the cold atom velocity and the distance $d$ that atoms travel out of the MOT. Under the radiation pressure from the pushing laser beam, the velocity is affected by the power and detuning and varies as
\begin{equation}
	\dfrac{{\rm d}v_z(t)}{{\rm d}t}=\dfrac{\hbar k \Gamma}{2m}\dfrac{s}{1+s+4(\delta-\vec{k}\cdot\vec{v}_z(t))^2/\Gamma^2},
\end{equation}
\noindent where $k$ is the wave vector, $\Gamma=2\pi\cdot5.22$ MHz the natural linewidth, $\delta=\omega_L-\omega_a$ the pushing laser detuning, and $s=I/I_s$ the saturation index. And the velocity and $t_{out}$ should meet the boudnary condition $d=\int_0^{t_{out}}v_z(t){\rm d}t$. In this way, the outcoupling rate $\Gamma_{out}$ can be calculated. 

With the balance of loading and loss, the rate equation of the 2D-HP MOT can be written as
\begin{equation}
R(n,v_z)-\Gamma_{trap}N-\Gamma_{out}N-\beta_N\dfrac{N^2}{V}=0,
\end{equation}
\noindent where $N$ is the cold atom number within the MOT. On the left side of Eq. (3), the third item, $\Gamma_{out}N$, represents the cold atomic flux of a certain axial velocity. It can be expressed as
\begin{eqnarray}
\Phi(n,v_z)&=&\Gamma_{out}(\dfrac{-V(\Gamma_{out}+\Gamma_{trap})}{2\beta_N}\nonumber\\
&+&\dfrac{\sqrt{V^2(\Gamma_{out}+\Gamma_{trap})^2+4V\beta_NR(n,v_z)}}{2\beta_N})\nonumber\\
&\times&\dfrac{v_z}{\Gamma_{coll}}(1-{\rm exp}(\dfrac{\Gamma_{coll}L}{v_z})),
\end{eqnarray}
\noindent where $\Gamma_{coll}$ is the loss rate due to the light-assisted collisions between the atomic beam and the thermal vapor. The value of $\Gamma_{coll}$ is proportional to the density of the background vapor, and usually one-order higher than $\Gamma_{trap}$ \cite{12J.Schoser}. The total atomic flux of the 2D-HP MOT can be obtained by the integral Eq. (4) over all axial velocity $\Phi_{tot}=\int_0^{\infty}\Phi(n,v_z){\rm d}v_z$. 
\begin{figure}[h]
	\includegraphics[width=\linewidth]{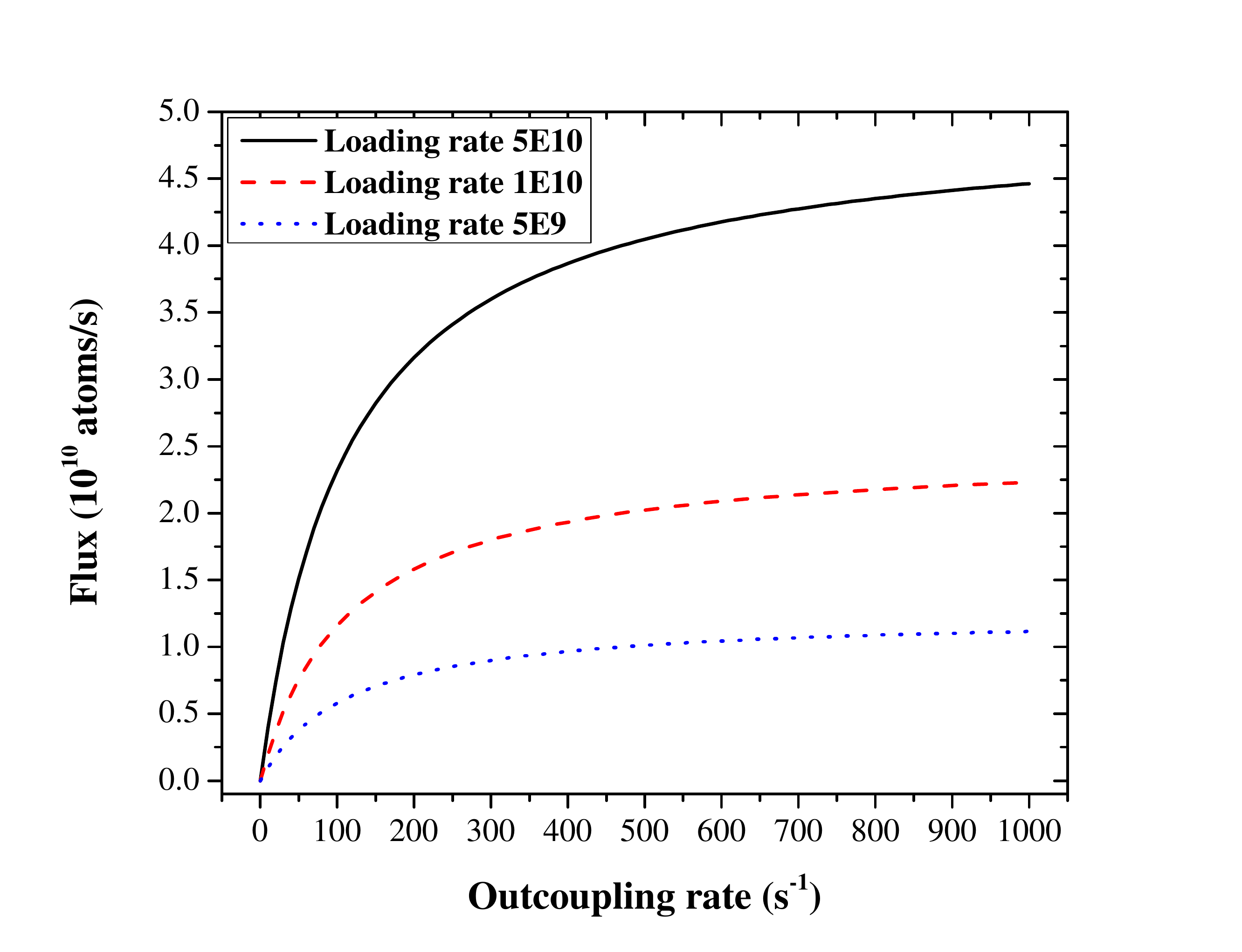}
	\caption{\label{fig:wide}The dependence of the total atomic flux on the outcoupling rate at different loading rates: $5\times 10^{10}$ atoms/s (black solid line), $1\times 10^{10}$ atoms/s (red dash line) and $5\times 10^9$ atoms/s (blue dot line).}
\end{figure}

According to Eq. (4), the dependence of the total atomic flux on the outcoupling rate can be calculated. The results calculated with different loading rates, $5\times 10^{10}$ atoms/s, $1\times 10^{10}$ atoms/s and $5\times 10^{9}$ atoms/s are as shown in Fig. 1. A high loading rate results in an intense flux. And the total atomic flux increases with the outcoupling rate until saturated. Thus, to achieve an intense cold atomic beam, both the loading rate and the outcoupling rate should be adjusted to high value.

Taking the average distance as $d=L/2=20$ mm and the initial axial velocity $v_z(0)=0$, the outcoupling rate $\Gamma_{out}$ to different pushing power and detuning (pushing beam diameter=3 mm) can be calculated with Eq. (2) and the boundary condition. And with the loading rate $5\times 10^{10}$ atoms/s, the dependences of the atomic flux on the pushing power and detuning are then obtained. As shown in Fig. 2(a), the atomic flux increases with the pushing power until saturated. The resonant pushing beam generates the highest flux. With the same detuning range, the red-detuned beam performs better than the blue-detuned one. And in Fig. 2(b), we can find that the atomic flux decreases when the pushing laser detuning increases. As the pushing power grows, the laser frequency of the peak flux red-shifts, and the flux distribution boradens. These flux variations of the zero-velocity atoms provide a pratical guidance for the pushing optimization in the 2D-HP MOT.
\begin{figure}[h]
	\includegraphics[width=\linewidth]{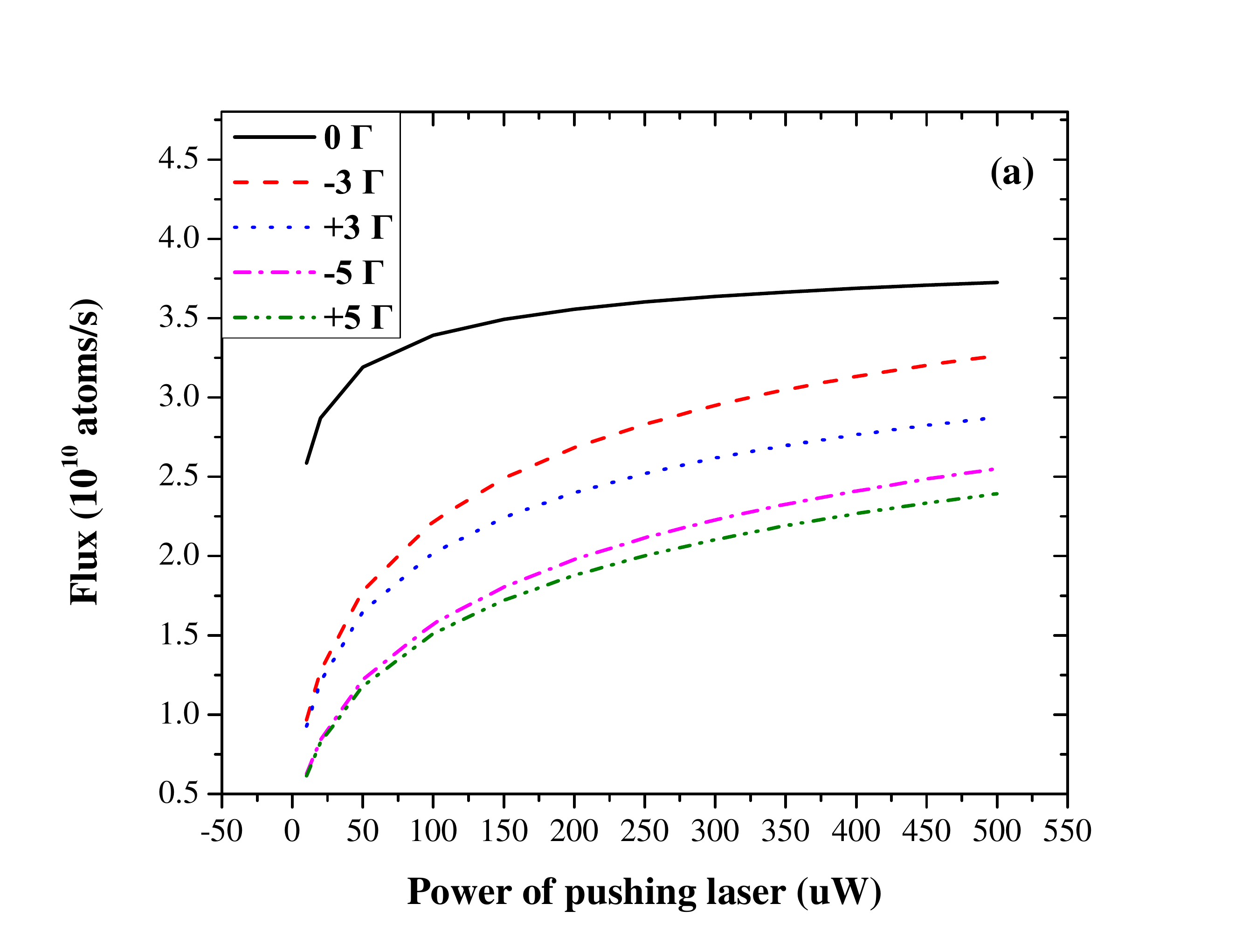}
	\includegraphics[width=\linewidth]{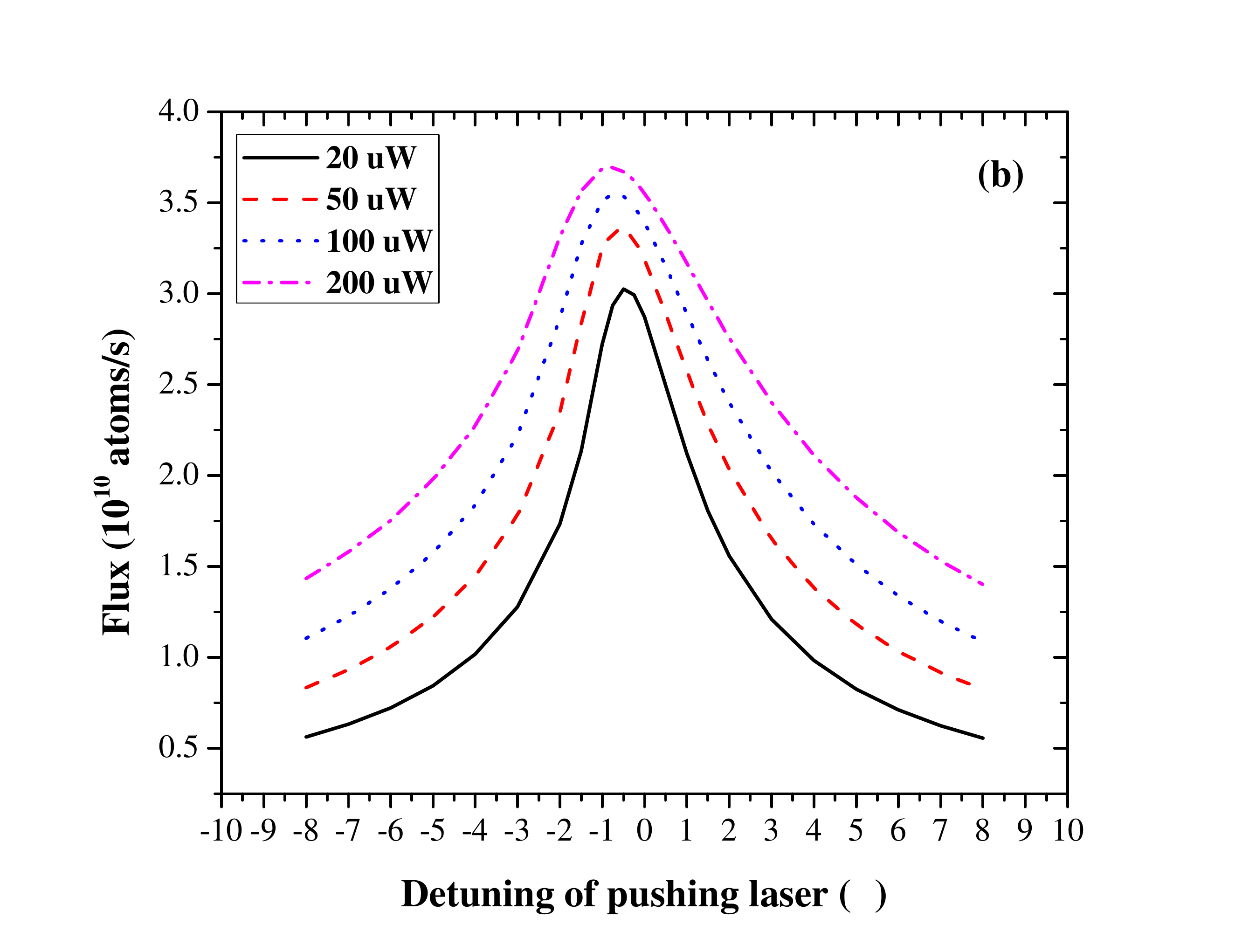}
	\caption{\label{fig:wide}The calculated atomic flux of the zero-velocity atoms: (a) the flux varying with the pushing power of various detunings, $0\Gamma$ (black solid line), $-3\Gamma$ (red dash line), $+3\Gamma$ (blue dot line), $-5\Gamma$ (magenta dash dot line), and $+5\Gamma$ (green dash dot dot line); (b) the flux varying with the pushing detuning of various power, $20 \mu W$ (black solid line), $50 \mu W$ (red dash line), $100 \mu W$ (blue dot line) and $200 \mu W$ (magenta dash dot line).}
\end{figure}

In addition to the independently-optimized outcoupling rate, the 2D-HP MOT can help suppress the light shift or other damaging effects generating by the leaking laser light from the aperture. The light shift $\Delta\omega_F$ to the cesium atom is \cite{25Affolderbach}
\begin{equation}
\Delta\omega_F=\dfrac{1}{4}\sum_{F\rq=F-1}^{F+1}\left|{\Omega_{FF\rq}}\right|^2\dfrac{\delta_{FF\rq}-\vec{k}\cdot\vec{v}_z}{(\delta_{FF\rq}-\vec{k}\cdot\vec{v}_z)^2+\Gamma^2/4},
\end{equation}
\noindent where $\Omega_{FF\rq}$ is the Rabi frequency, $\left|{\Omega_{FF\rq}}\right|^2=\Gamma^2I/2I_s$ is proportional to the laser intensity, $\delta_{FF\rq}=\omega_L-\omega_{FF\rq}$ is the laser detuning, and $F=$3 or 4 correspond to the ground states. As the thin pushing beam weakening the leaking light, the light shift would be effectively suppressed. Moreover, the fluctuation of the light shift, which is much more destructive to signal contrast, can also be significantly suppressed. 

\begin{figure*}\centering
	\includegraphics[width=\textwidth]{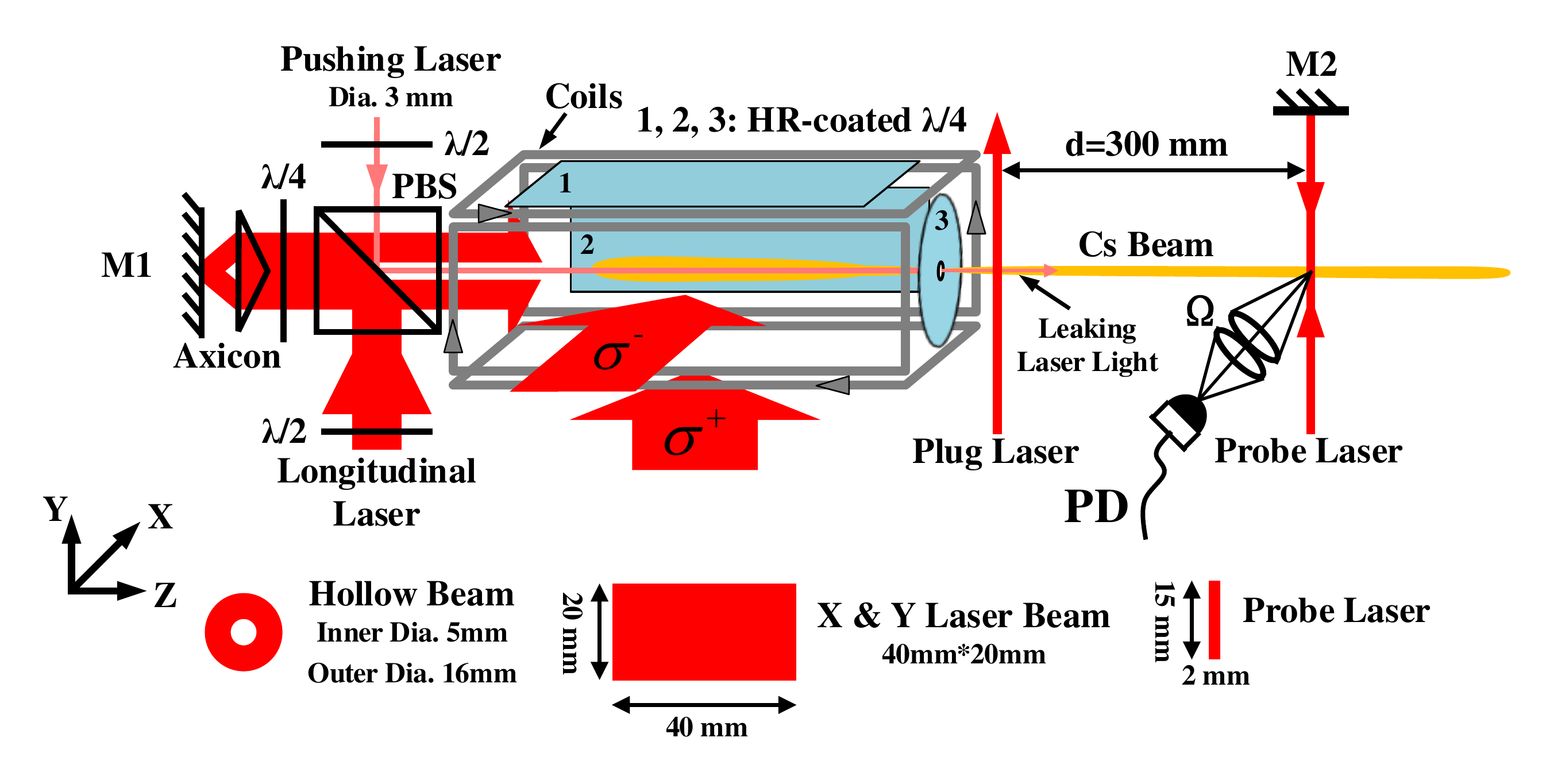}
	\caption{\label{fig:wide}The schematic diagram of the 2D-HP MOT. M$_1$, M$_2$: mirror; $\lambda$/4, $\lambda$/2: wave plate; PBS: polarized beam splitter; HR-coated $\lambda$/4: $\lambda$/4 wave plate coated with high-reflectivity film; $\sigma^+$, $\sigma^-$: circular polarization; $\Omega$: collection angle; PD: photodiode.}	
\end{figure*}

\section{EXPERIMENTAL SETUP AND DIAGNOSTICS}

The experimental setup consists of the vacuum system, the quadrupole magnetic field, and the laser system. The cesium vapor cell is formed by a 55 mm-diameter, long circular quartz tube. Both ends of the tube are connected to the pump system. And near the detection region, an ion pump is additionally employed to keep a higher vacuum. A temperature-controlled cesium reservoir is connected to the tube beside the MOT via a vacuum valve. In the experiments, the vacuum is $1.2\times 10^{-5}$ Pa at the MOT, and $3.5\times 10^{-6}$ Pa at the detection region.

The quadrupole magnetic field is generated by two orthogonal pairs of $140\times 100$ mm rectangular coils, as shown in Fig. 3. The coils are spaced by 110 mm and symmetric in XY directions. Along the Z direction, it retains no magnetic field. The XY magnetic field gradient is set to 10 G/cm.

The diode laser system includes a DBR laser, a DFB laser and an extended cavity diode laser (ECDL), which are all frequency-locked using the standard saturation absorption (SA) techniques. The DBR laser beam is split into two parts: one major part is used for the cooling and the other part is used as the pushing beam. The cooling frequency is detuned $-2.5\Gamma$ from the cycling transition $6^2S_{1/2} (F=4)\to 6^2P_{3/2} (F\rq=5)$. By means of two acoustic optical modulator (AOM), the pushing frequency can be shifted from $-8\Gamma$ to $8\Gamma$ around the $4\to 5\rq$ cycling transition. The DFB laser serves as the repumping laser and is frequency-locked to the transition $6^2S_{1/2} (F=3)\to 6^2P_{3/2} (F\rq=4)$. The probe laser and the plug laser are derived from the ECDL laser beam. And the plug laser is resonant with the $4\to 5\rq$ cycling transition. 

The configurations of the laser beams are presented in Fig. 3. To obtain a large cooling volume, the XY laser beams are expanded into 40 mm$\times$20 mm rectangles by cylindrical lens. Containing 27-mW cooling laser and 4-mW repumping laser, the XY laser beams are circularly polarized and retroreflected by $\lambda$/4 wave plates (50 mm$\times$25 mm). The $\lambda$/4 wave plates are coated with high-reflectivity film on the back to make the laser beams couterpropagate and polarized reversely. In the axial direction, a hollow beam, which is linearly polarized and retroreflected by another $\lambda$/4 wave plate, is applied for cooling as designed. The hollow beam is generated by an axicon, with the inner diameter 5 mm and the outer diameter 16 mm. And the power is 15 mW. At the center of the wave plate, a 1-mm aperture is drilled for atomic beam extraction. The pushing laser beam (beam diameter=3 mm) is combined with the hollow beam by a polarized beam splitter (PBS). The pushing beam, the hollow beam and the aperture are aligned concentrically. In the experiment, a part of the pushing laser would leak out of the aperture and go along with the atomic beam. Here it is marked as ``leaking laser light.''

The plug laser and the probe laser are used for diagnosing the cold atomic beam. The plug laser is 2.5 mW with a 5-mm diameter, and the probe laser is a 200-$\mu$W rectangular beam (15 mm$\times$2 mm). The probe laser beam and plug laser beam are sparated by 300 mm. To enhance the fluorescence signal, the probe laser beam is retroreflected. The fluorescence emitted from the cold cesium atoms is collected by the lens assembly with a spatial angle $\Omega$, and measured by a photodiode (Hamamatsu S3204-08). A time-of-flight (TOF) method is used to measure the atomic beam flux and the axial velocity distribution. Suddenly opening the plug beam, the cold atoms would be pushed away the axis. And the time dependence of the decaying fluorescence signal $S(\tau)$ is detected. Then the axial velocity distribution $\Phi(v_z)$ can be deduced as
\begin{equation}
\Phi(v_z)=-\dfrac{\eta}{d_{probe}}\dfrac{l}{v_z}\dfrac{{\rm d}S(\tau)}{{\rm d}\tau},
\end{equation}
\noindent where $\eta$ is a calibration factor of the detection system, $d_{probe}=2$ mm is the width of the probe laser beam, and $l=300$ mm is the distance between the plug laser and the probe laser. The total atomic flux in the experiments can be obtained by an over all integral of Eq. (6). 

\section{EXPERIMENTAL RESULTS AND DISCUSSIONS}

In experiments, the 2D-HP MOT is realized, where the axial cooling and pushing are accomplished by a hollow beam and another pushing beam, respectively. The dependences of the total atomic flux on the pushing power and detuning are measured, and a comparison is made between the 2D-HP MOT and the traditional 2D$^+$ MOT based on the experimental results.

\subsection{Demonstration for the 2D-HP MOT}

In the 2D-HP MOT, the hollow beam keeps the loading rate but makes no contribution to the atomic beam extraction, and the pushing beam regulating the outcoupling rate determines the atomic beam extraction. In experiment, the pushing beam is 15 $\mu$W and resonant with the $4\to 5\rq$ cycling transition. By sweeping the probe laser frequency around the $4\to 5\rq$ cycling transition, the hyperfine spectral lines of the cold atomic beam in different conditions are measured as shown in Fig. 4(a). When only the hollow beam is on, the cold cesium beam is not extracted, and the spectral line (blue dot line) is almost invisible. When only the pushing beam is on, the spectral line (the red dash line) is still very weak, because the loading rate is rather low without the axial cooling. And when both the hollow and pushing beams are on, the spectral line (black solid line) becomes very distinct. And the signal is enhanced by one order of magnitude than that in the only-pushing condition.

Besides, the axial velocity distribution in the pushing-only condition and the cooling-pushing condition is measured, as shown in Fig. 4(b). When only the pushing laser is on (red dash line), the mean velocity of the distribution is 7.6 m/s and the total flux is $4.18\times 10^9$ atoms/s. When both the cooling and pushing are on (black solid line), the mean velocity is reduced to 7.0 m/s, and the total flux is enhanced to $3.03\times 10^{10}$ atoms/s. The flux increases by about one order of magnitude, just as that of the spectral lines.

These results demonstrates that the 2D-HP MOT works as designed. The hollow beam keeps the loading rate and the pushing beam determines the outcoupling rate. An intense cold atomic beam is extracted when both laser beams are on. 
\begin{figure}[h]
	\includegraphics[width=\linewidth]{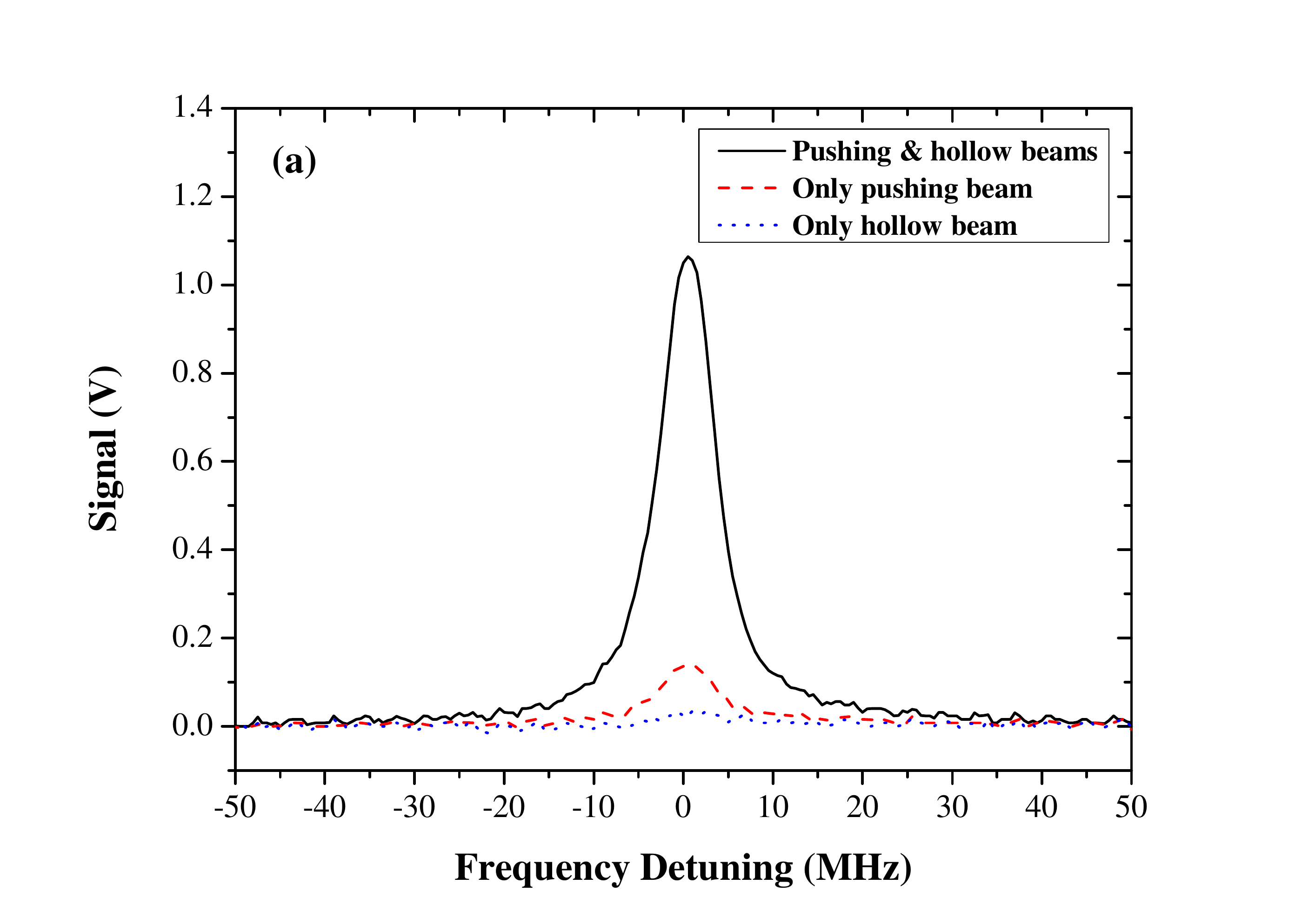}
	\includegraphics[width=\linewidth]{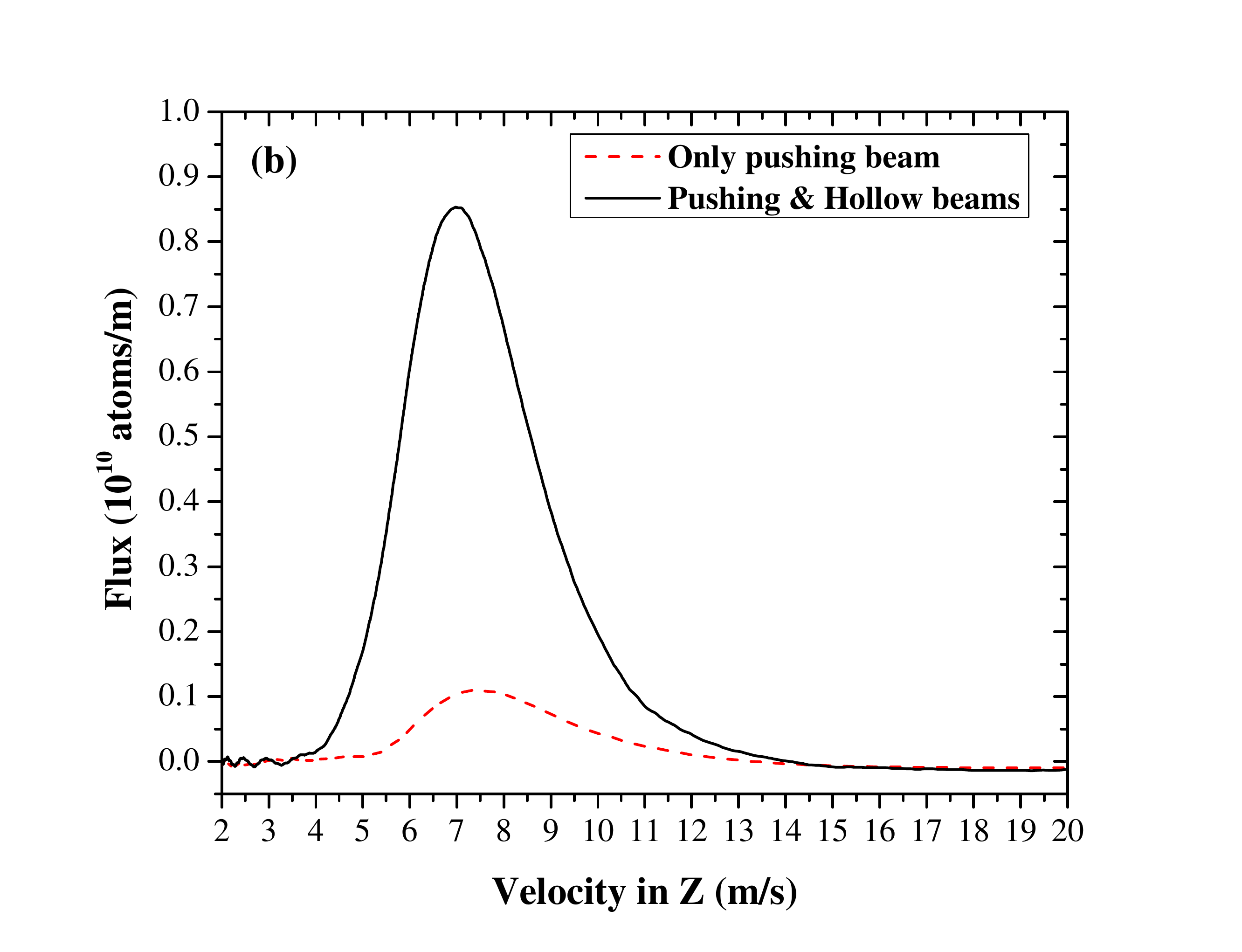}
	\caption{\label{fig:wide}Signal of the cold cesium atomic beam in different conditions: (a) the $4-5\rq$ hyperfine spectral lines: both hollow and pushing beams on (black solid line), only pushing beam on (red dash line), only hollow beam on (blue dot line); (b) the axial velocity distribution: only pushing beam on (red dash line), both hollow and pushing beam on (black solid line).}
\end{figure}

\subsection{Dependences on pushing parameters}

In experiments, the dependence of the total atomic flux on the pushing power is measured at the detuning of $-7\Gamma$, $-5\Gamma$, $0\Gamma$, $+5\Gamma$ and $+7\Gamma$, as shown in Fig. 5(a). The variations present that the total atomic flux would increase with the pushing power until saturated. This incensement is caused by the enhanced outcoupling rate. At zero-detuning condition, a weak pushing power leads to a relatively high flux. And the atomic flux reaches the saturation value, $3.03\times 10^{10}$ atoms/s, when the power is only 15 $\mu$W. The atomic flux at red detuning is higher than that at blue detuning. The maximum flux in experiments, $4.02\times 10^{10}$ atoms/s, is obtained when the pushing power is 400 $\mu$W and the detuning is at $-5\Gamma$. These results are in agreement with the theoretical calculation in Fig. 2(a). 
\begin{figure}[h]
	\includegraphics[width=\linewidth]{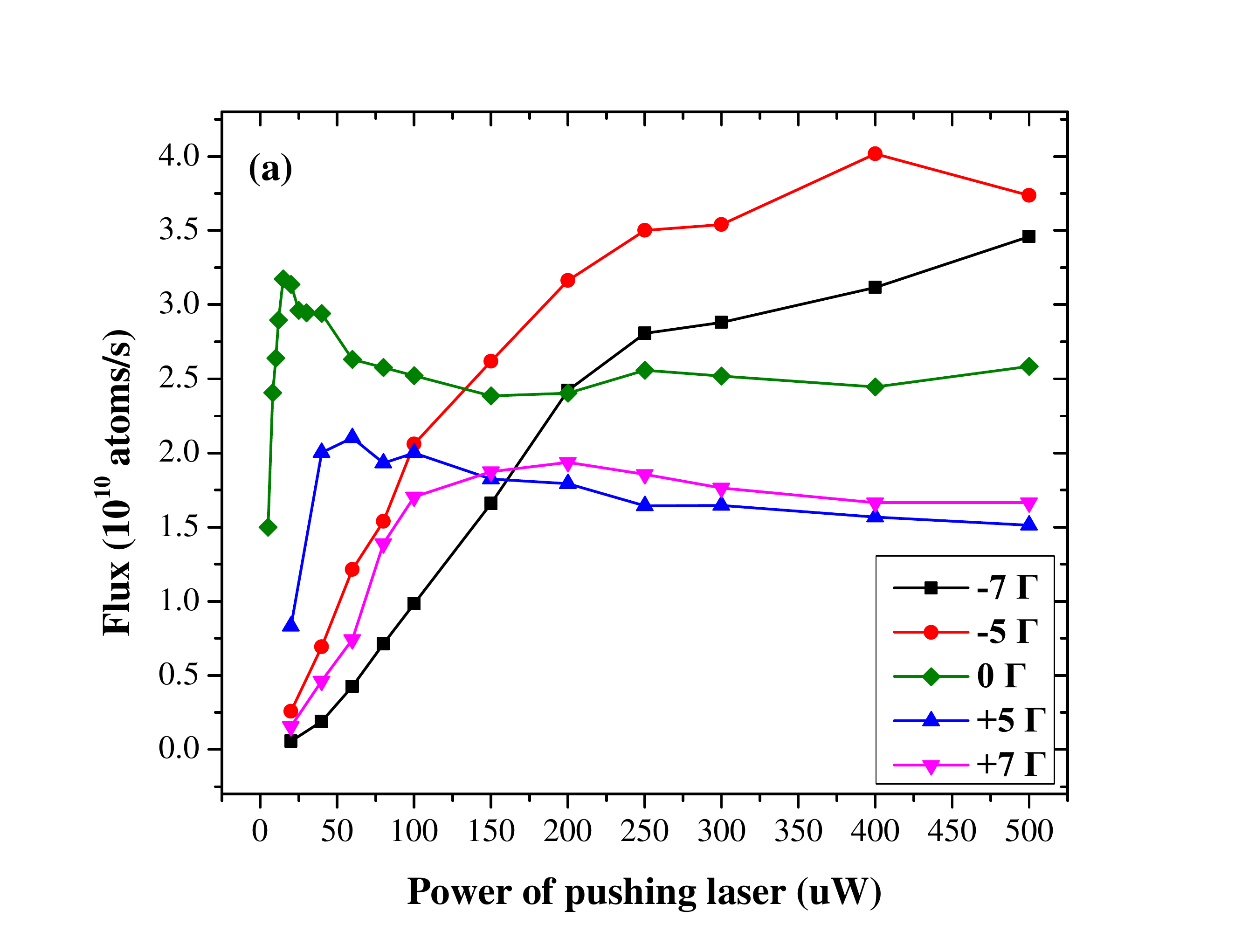}
	\includegraphics[width=\linewidth]{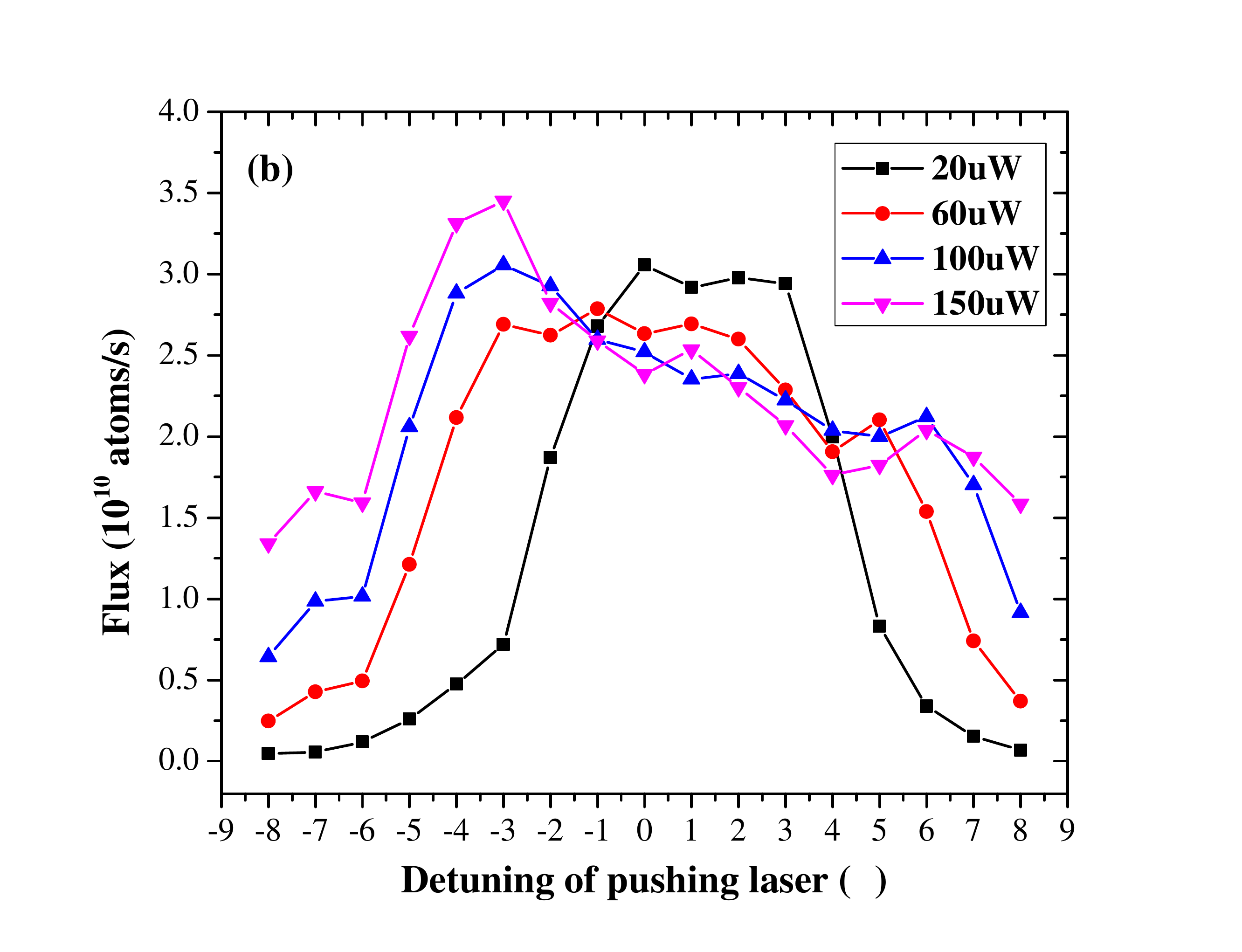}
	\caption{\label{fig:wide}The dependences of the total atomic flux on the pushing parameters: (a) the dependence on the pushing power at different pushing detuning: $-7\Gamma$ (black square line), $-5\Gamma$ (red circle line), $0\Gamma$ (green diamond line), $+5\Gamma$ (blue up triangle line) and $+7\Gamma$ (magenta down triangle line); (b) the dependence on the pushing detuning at different pushing power: 20 $\mu$W (black square line), 60 $\mu$W (red circle line), 100 $\mu$W (blue up triangle line) and 150 $\mu$W (magenta down triangle line).}
\end{figure}

However, the experimental results in Fig. 5(a) also show one condition which is not predicted in theoretical calculations. When the pushing power increases to a certain value, the atomic flux begins to decrease. The turning point of laser power has the minimum value at zero-detuning condition, and increases as the pushing detuning extends. The turning power at red-detuning condistions has a higher value than that at blue-detuning conditions. 

We consider that the decline of the loading rate caused by the heating effect around the axis mainly accounts for the flux loss. According to the calculation results in Fig. 1, the atomic flux is sensitive to the loading rate. And in experiments, the pushing laser beam beyond the aperture can generate a heating effect to cold cesium atoms around the axis, which is not included in Eq. (4). The heating effect impacts the axial cooling and directly results in a loading rate decline. Hence, even though the pushing laser provides a rather high outcoupling rate, the atomic flux would decrease when the pushing power is beyond a certain value. The heating effect will become obvious when the pushing laser is intense or close to rensonance. And the blue-detuned laser heats the cold cesium atoms more serious than the one red-detuned.

Additionally, the dependence of the total atomic flux on the pushing detuning is measured at pushing power of 20 $\mu$W, 60 $\mu$W, 100 $\mu$W and 150 $\mu$W are measured as shown in Fig. 5(b). We can find that the total atomic flux has a higher value when the frequency of the pushing laser is close to resonance. And with the pushing power increasing, the laser frequency corresponding to the peak flux red-shifts, and the high-flux range broadens. These variation are consistent with the theoretical calculations in Fig. 2(b). Besides, the atomic flux measured in Fig. 5(b) correlates linearly with the pushing detuning within a frequency range. This phenomenon is interesting but not revealed in Fig. 2(b), because the calcualtion results of zero-velocity atoms cannot fully reflect the reality. We consider that the combined effects of the pushing beam on the cold cesium atoms following the maxwell distribution accounts for the phenonmenon.

\subsection{Comparison with the 2D$^+$ MOT}

To compare to the 2D$^+$ MOT, the 2D-HP MOT is modified by blocking the pushing beam and replacing the hollow laser beam by a regular Gaussian beam. The laser beam has the same diameter, power and detuning as the previous hollow beam. The central intensity of the Gaussian beam is the pushing intensity, 8.06 mW/cm$^2$. And other parts of the setup in Fig. 3 remains unchanged. Thus, except the fixed pushing, other parameters of the 2D$^+$ MOT the same as the 2D-HP MOT. 

The axial velocity distribution of the cold atomic beam extracted from the 2D$^+$ MOT centers at 7.2 m/s with a FWHM of 3.2 m/s, the black solid line shown in Fig. 6. The total atomic flux is $2.54\times 10^{10}$ atoms/s. 

To make a comparison with the 2D$^+$ MOT, the velocity distributions obtained in the 2D-HP MOT with the pushing parameters (-5$\Gamma$, 400 $\mu$W) and (0$\Gamma$, 10 $\mu$W) are also plotted in Fig. 6. For the parameters (-5$\Gamma$, 400 $\mu$W), the pushing laser intensity is 5.66 mW/cm$^2$. In this condition, the total atomic flux reaches the experimental maximum, $4.02\times 10^{10}$. And the velocity distribution centers at 8.2 m/s with a FWHM of 4.2 m/s (the blue dot line in Fig. 6). Compared to the 2D$^+$ MOT, the 2D-HP MOT with proper pushing can enhance the atomic flux by 60 percent. This enhancement would boost the SNR of the experiments using cold cesium beam, and accelerate the UHV MOT loading. 
\begin{figure}[h]
	\includegraphics[width=\linewidth]{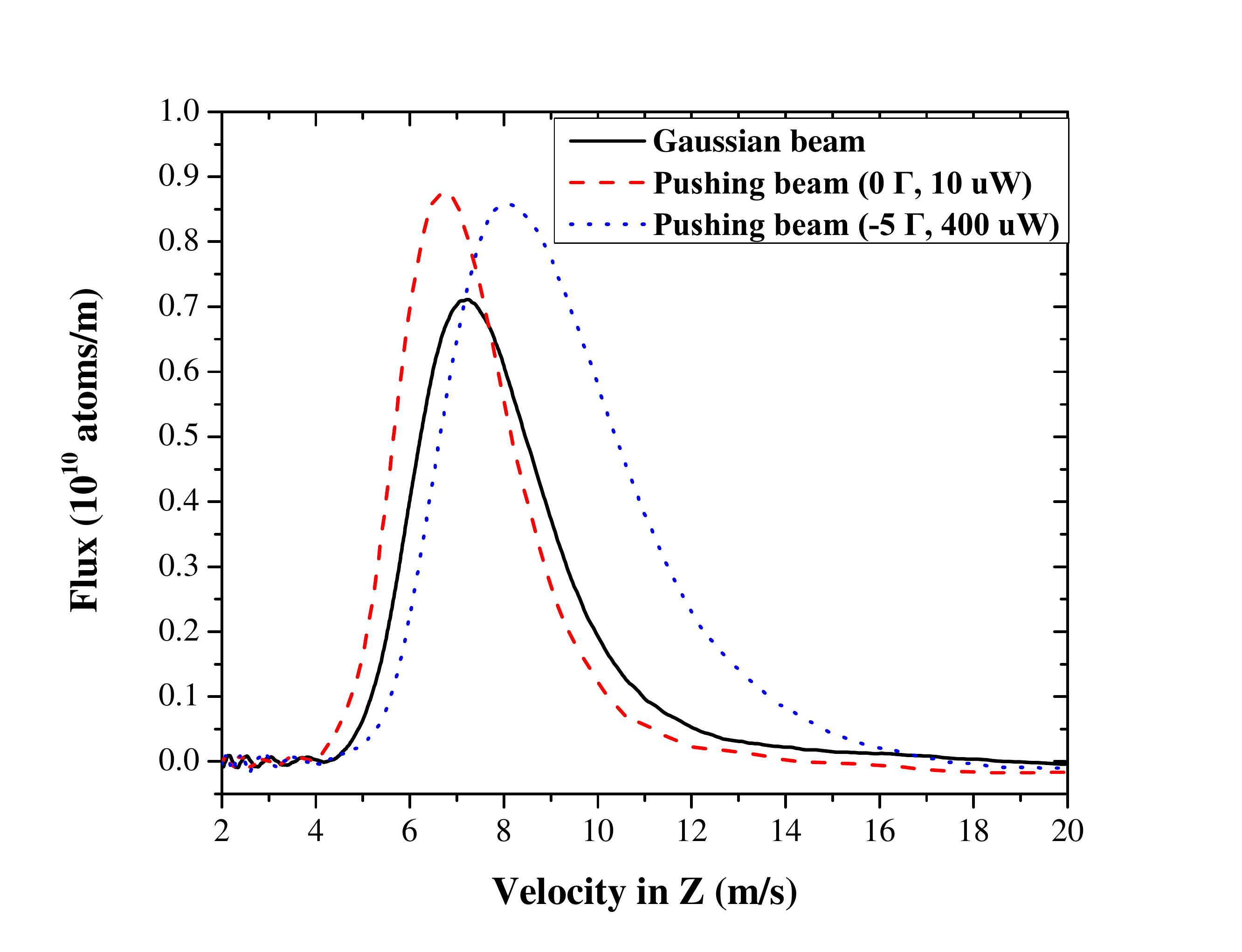}
	\caption{\label{fig:wide}The axial velocity distributions of the cold atomic beams: 2D$^+$ MOT (solid beam, black solid line), 2D-HP MOT (0$\Gamma$, 10 $\mu$W, red dash line), and 2D-HP MOT (-5$\Gamma$, 400 $\mu$W, blue dot line).}
\end{figure}

For the parameters (0$\Gamma$, 10 $\mu$W), the pushing laser intensity is 0.14 mW/cm$^2$. In this condition, the atomic flux is $2.64\times 10^{10}$ atoms/s. And the velocity distributes around 6.8 m/s with a FWHM of 2.8 m/s (the red dash line shown in Fig. 6). Compared to the 2D$^+$ MOT, the 2D-HP MOT of this condition generates similar flux but consumes much less pushing power. This thin pushing beam can help effectively suppress the light shift causing by the leaking laser light. According to Eq. (5), the light shift of the ground state hyperfine transition between $6^2S_{1/2} (F=3)$ and $6^2S_{1/2} (F=4)$ caused by the thin pushing laser is -19.76 kHz, while the one by the solid beam -436.82 kHz. The light shift in the 2D-HP MOT is about 20 times smaller than that in the 2D$^+$ MOT. Moreover, the fluctuation of the light shift due to the power instability will be drastically reduced with the same relative laser power fluctuation. This suppression to the light shift fluctuation can make great improvement in the signal contrast of the interference signal in atomic clocks or atom interferometers. 

The 2D-HP MOT needs some more facilities than the 2D$^+$ MOT to generate the hollow beam and the pushing beam. However, with proper pushing parameters, the 2D-HP MOT can generate a relatively intense cold atomic beam with rather low concomitant light shift, which would be much more desirable than that of the 2D$^+$ MOT in precision spectroscopy applications.

\section{CONCLUSION AND OUTLOOK}

In this work, we present a new desgin for source of cold cesium atoms, the 2D-HP MOT. With independent axial cooling and pushing, the 2D-HP MOT can substantially optimized the atomic flux. The atomic flux maximum obtained in the 2D-HP MOT is $4.02\times 10^{10}$ atoms/s, 60 percent higher than the traditional 2D$^+$ MOT. And with proper pushing parameters, the 2D-HP MOT can generate a rather intense cold atomic beam with the concomitant light shift 20 times smaller than that in the 2D$^+$ MOT. Thus, compared to the traditional 2D$^+$ MOT, the 2D-HP MOT provides much more valuable benefits for precision measurement experiments, for example, an atomic clock based on a continuous cold cesium atomic beam. 

\begin{acknowledgements}
	This work was supported by the National Natural Science Foundation of China (11304177). The authors specially acknowledge Lu Zhao, Chi Xu, Wei Xiong and Yu-Hang Li for their helpful discussions.
\end{acknowledgements}

\bibliography{2dhpmot}% Produces the bibliography via BibTeX.

\end{document}